# Bulk Topological Superconductors

SHINGO YONEZAWA

DEPARTMENT OF PHYSICS, GRADUATE SCHOOL OF SCIENCE, KYOTO UNIVERSITY

**ABSTRACT**

This review introduces known candidates for bulk topological superconductors and categorizes them with time-reversal symmetry (TRS) and gap structures. Recent studies on two archetypal topological superconductors, TRS-broken $Sr_2RuO_4$ and TRS-preserved $Cu_xBi_2Se_3$, are described in some detail.

## 1. INTRODUCTION

In recent years, research on topological materials has been explosively growing. This explosion was triggered by the prediction and discovery of quantum spin-Hall systems and topological insulators (TIs), with non-trivial topology of the valence-band wave function in the reciprocal space. Similarly, it has been recognized that some fully-gapped superconductors also possess non-trivial topological properties in their superconducting (SC) wave functions [1-6]. Such superconductors are called topological superconductors (TSCs).

One of the most intriguing properties of gapped topological materials is bulk-edge correspondence. Because it is by definition impossible to change a topologically non-trivial electronic state to a trivial state by a continuous deformation, there must be a certain kind of "singularity" at the boundaries between topologically distinct materials. Such considerations explain the closing of the gap, i.e. existence of gapless states, on the boundary between a gapped topological material and a gapped non-topological material (including vacuums).

Fully-gapped topological superconductivity is also accompanied by gapless surface states. Furthermore, because of the particle-hole symmetry in the low-energy Hamiltonian, surface states of a TSC should have the Majorana nature, i.e., particles being identical to their own antiparticles [6]. Intriguingly, the surface state at zero energy (i.e., at the chemical potential), called the Majorana zero modes, do not obey ordinary (Fermionic or Bosonic) statistics in two dimensions. Instead, the Majorana zero modes exhibit non-Abelian statistics, which can be potentially utilized for quantum computing [7].

Although the concept of topology was originally considered for systems with full energy gaps, it has recently expanded to gapless (i.e. gaps with nodes) states [5]; one can evaluate a "topological invariant of a gap node" on a manifold surrounding the node. Similarly, on a certain cut of the Brillouin zone, a $k$-dependent topological number can be defined. Application of this extension to superconductivity gives us new insights on unconventional superconductors.

## 2. TOPOLOGICAL SUPERCONDUCTIVITY – CANDIDATES FOR MATERIALS

Within the conservative definition based on global topological numbers in fully gapped systems, materials exhibiting topological superconductivity are rather scarce. Nevertheless, there are several known candidates for bulk TSCs and superfluids as listed in Table I. With the recent expansion of the topology concept to nodal superconductors, a number of nodal unconventional superconductors studied over 35 years now can be added to the list, although topological phenomena in most of the new members have not been experimentally explored. Gap structures of representative topological SC and superfluid states are schematically shown in Fig. 1.

**Prototypical superfluid $^3$He phases**
The most classical and established cases are the superfluid phases in $^3$He. Both the A and B phases of the superfluid $^3$He have spin-triplet odd-parity Cooper pairs [8]. The former phase, the 3D chiral $p$-wave state, spontaneously *breaks* TRS and is topologically non-trivial because of topological monopole-like point nodes (Weyl nodes). The B phase, in which TRS is *preserved*, is also topologically non-trivial, because of the hedgehog-like spin-orbital texture on its Fermi surface. The $^3$He phases have been providing prototypical playgrounds to investigate properties of topological pairings [9,10].

| Time-reversal symmetry | Gap nodes | Material | Possible pairing state | Surface states | Remarks |
|---|---|---|---|---|---|
| Spontaneously Broken | Full gap | $Sr_2RuO_4$ | Q2D chiral-$p$ | Observed | • Tetragonal<br>• Deep gap minima? |
| | | SrPtAs | Q2D? chiral-$d$ | | • Hexagonal |
| | Nodal gap | $^3$He; A phase | 3D chiral-$p$ (ABM state) | | • Superfluid<br>• Weyl superfluid |
| | | UPt$_3$; B phase | 3D chiral-$f$ | | • Heavy fermion, trigonal<br>• Weyl superconductor |
| | | URu$_2$Si$_2$ | 3D chiral-$d$ | | • Heavy fermion, hidden order, tetragonal<br>• Weyl superconductor |
| Preserved | Full gap | $^3$He; B phase | 3D $p$ (BW state) | Observed | • Superfluid |
| | | $A_x$Bi$_2$Se$_3$ ($A$ = Cu, Sr, Nb, Tl) | Q2D (3D?) $p$ or $f$ | Observed | • Doped TI, hexagonal |
| | | Sn$_{1-x}$In$_x$Te | 3D | Observed | • Doped TCI, cubic |
| | Nodal gap | High-$T_c$ cuprates | Q2D $d_{x2-y2}$ | Observed | • Tetragonal (orthorhombic for some cases) |
| | | Ce$T$In$_5$ ($T$ = Co, Rh, Ir) | Q2D $d_{x2-y2}$ | | • Heavy fermions, tetragonal |
| | | $\kappa$-(ET)$_2$X | Q2D $d$-wave-like | | • Organic, monoclinic etc. |
| | | (TMTSF)$_2$X | Q1D $d$-wave-like | | • Organic, triclinic |
| | | Cu$_x$(PbSe)$_5$(Bi$_2$Se$_3$)$_6$ | Q2D nodal | | • Doped TI/BI natural superlattice, |
| | | CePt$_3$Si, Ce$T$Si$_3$ ($T$ = Rh, Ir) | Q2D $p + s$ | | • Non-centrosymmetric, heavy Fermion, tetragonal |
| | | Li$_2$Pt$_3$B | 3D $p + s$ | | • Non-centrosymmetric, cubic |
| Already broken in the normal state | ? | UGe$_2$, URhGe, UCoGe | Non-unitary $p$ or $f$ | | • Ferromagnetic superconductors |

**Table I.** List of known candidate materials for bulk topological superconductors and superfluids.

**Classification based on TRS and gap nodes**
Here, I classify bulk topological pairing states based on accompanying TRS breaking as shown in Table I. Each class is then divided into states with full gaps and those with nodal gaps. I believe that this way is convenient to grasp the current experimental situations, because TRS breaking and gap structures can be clearly detectable by experiments and indeed, have been investigated in most of the candidate materials.

**TRS-*broken* topological superconductivity**
Topological SC states with spontaneous TRS breaking due to orbital phase winding can be termed as "chiral" states. The layered perovskite oxide Sr$_2$RuO$_4$ is a leading candidate for quasi-two-dimensional (Q2D) chiral-$p$-wave superconductors [11-14]. The current research situation on this superconductor is described in §3. In the recently-discovered superconductor SrPtAs with a hexagonal honeycomb structure, spontaneous TRS breaking was reported based on μSR experiments, suggesting possible chiral-$d$-wave states [15]. A fully-gapped state is indicated by the NMR study [16]. However, the observed coherence peak in $1/T_1$ may favor $s$-wave superconductivity.

Among the nodal superconductors, two heavy-fermion superconductors, UPt$_3$ and URu$_2$Si$_2$, are leading candidates for TRS-broken topological superconductivity. TRS breaking in the B-phase of UPt$_3$ had been indicated 23 years ago by a μSR experiment [17] but with controversy [18], and has been recently confirmed by the Kerr effect study [19]. Since the odd-parity nature of superconductivity in this compound has been revealed by using the NMR technique [20] and polarized neutron scattering [21], the pairing in UPt$_3$ can be most promisingly interpreted as the chiral-$f$-wave $k_z(k_x \pm ik_y)^2$ or $(5k_z^2 − 1)(k_x ± ik_y)$ states (see Fig. 1), which are accompanied by topological Weyl nodes [22,23]. We note that some fundamental issues such as existence of tetra critical point have not been explained within the scenario of chiral states [22]. The heavy fermion URu$_2$Si$_2$ also exhibits TRS breaking in the SC state as evidenced by the Kerr effect [24]. Another signature for the chiral state is also obtained by the Nernst effect, which is found to be exceptionally huge just above $T_c$ of this compound [25] and is attributed to superconducting fluctuation with chiral nature [26]. Together with the Pauli-limit-like $H_{c2}$ behavior [27], it has been believed that superconductivity of this compound is of a 3D chiral-*d*-wave $k_z(k_x ± ik_y)$ (= $d_{zx} ± id_{yz}$) state (Fig. 1) with Weyl nodal points in its SC gap, although direct evidence for the spin-singlet SC nature is actually still controversial [28,29].

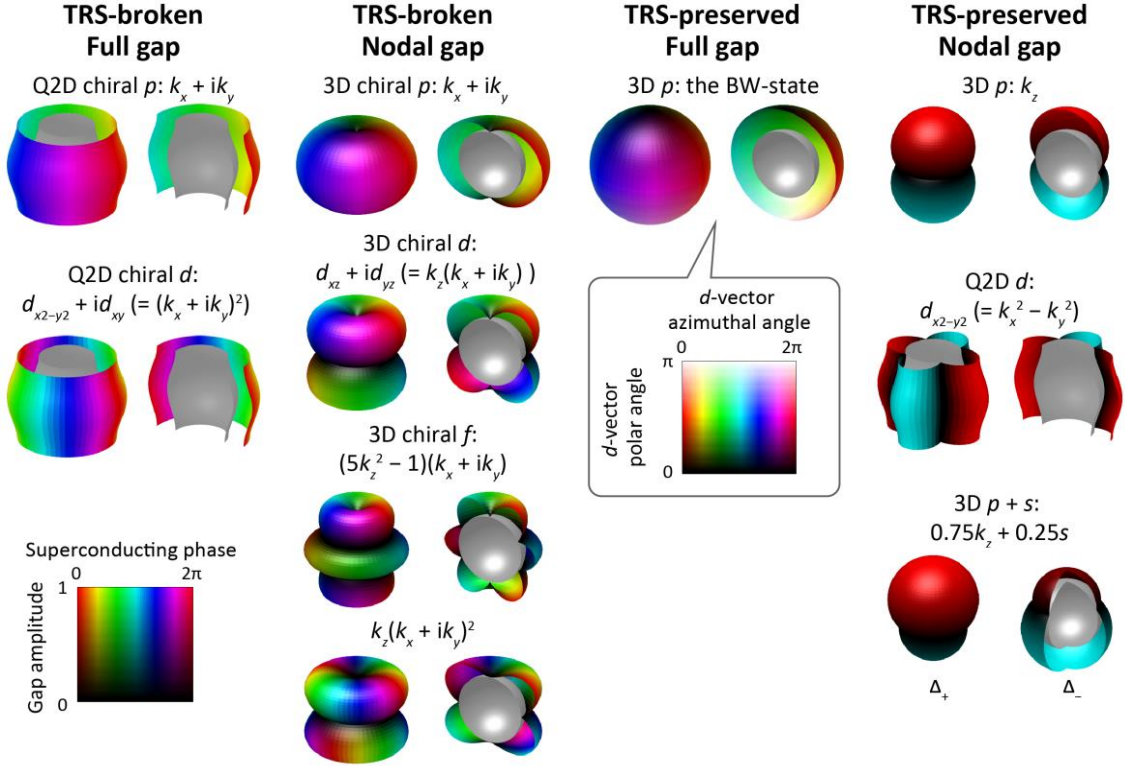

**Figure 1**. Schematics of representative bulk topological superconducting and superfluid states. The color depicts the superconducting gap amplitude and phase as shown in the left bottom panel, with black corresponding to gap nodes, except for the BW state (corresponding to the B phase of $^3$He), where the color represents the *d*-vector direction. For all structures, cross sectional views are also shown, in which the gray tubes or spheres indicate the normal-state Fermi surface. For the parity-mixed *p* + *s* state, positive and negative helicity gaps [5] are shown.

**TRS-*preserved* topological superconductivity**

In 2010, Hor *et al*. discovered that the topological insulator Bi$_2$Se$_3$ exhibits superconductivity below ~ 3.8 K when Cu ions are intercalated [30]. Theories and some experiments have indicated fully-gapped, topological, and odd-parity superconductivity in Cu$_x$Bi$_2$Se$_3$ [3,4], as described in more detail in §4. The doped topological crystalline insulator (TCI) Sn$_{1−x}$In$_x$Te was actually known to exhibit superconductivity below around 2 K [31], far before SnTe was recognized as a TCI. Point-contact spectroscopy in the SC state revealed the possibility of unusual surface states [32]. A fully-opening gap and preserved TRS are indicted by specific-heat [33] and μSR studies [34]. These results satisfy a condition for TRS-preserved fully-gapped topological SC states.

It should be noted that two heavy-fermion superconductors, namely the spin-singlet superconductor CeCu$_2$Si$_2$ and the spin-triplet superconductor UBe$_{13}$ (not listed in Table I), may have fully-opened gaps according to recent specific-heat studies [35,36], although they had been believed to be nodal superconductors. Although the gap functions are not clear, the possibility of fully-gapped topological superconductivity is worth investigating.

Candidates for TRS-preserved nodal TSCs are also listed in Table I. It has been established that high-$T_c$ cuprate superconductors exhibit Q2D $d_{x2-y2}$ superconductivity (Fig. 1) [37], and the nodes of those superconductors are protected due to a topological reason. Indeed, as early as 1995, surface states originating from the SC gap had been detected as a sharp zero-bias peak in tunneling spectroscopy [38], due to flat-band surface states originating from the non-trivial $k$-dependent topology [39]. Similarly, Q2D or Q1D $d$-wave or $d$-wave-like superconductors such as Ce$T$In$_5$ ($T$ = Co, Rh, Ir) [40], κ-(ET)$_2X$ ($X$ = Cu(NCS)$_2$ etc.) [41], and (TMTSF)$_2X$ ($X$ = ClO$_4$, etc.) [42] can be regarded as topological nodal superconductors. Recently, the Cu-doped (PbSe)$_5$(Bi$_2$Se$_3$)$_6$, a "naturally-made" superlattice between a TI and an ordinary insulator, is found to become a superconductor with $T_c$ ~ 3 K [43], probably with a line nodal SC gap.

Some superconductors with non-centro-symmetric crystal structures (i.e., structures without inversion symmetry) are also candidates for topological nodal superconductors [5]. In such superconductors, mixing of spin-triplet and spin-singlet order parameters are possible because of the lack of the parity in their crystal structures, and gap nodes emerge when the spin-triplet component is dominant (see Fig. 1) [44]. The tetragonal CePt$_3$Si and Ce$T$Si$_3$ ($T$ = Rh and Ir) exhibit large upper critical fields attributed to parity-mixed superconducting states [45-47]. In the cubic Li$_2$Pt$_3$B, spin-triplet-dominant and nodal superconductivity is reported based on NMR studies [48].

**Ferromagnetic superconductivity with broken TRS**
In the present classification, another interesting class is the ferromagnetic superconductors UGe$_2$, URhGe, and UCoGe [49], in which the TRS is *already broken in the normal state* due to the ferromagnetic ordering. In these superconductors, both superconductivity and ferromagnetism originate from electrons in the same (probably uranium) bands. Judging from this, as well as from the large upper critical fields, non-unitary (spin-polarized) spin-triplet states are likely to be realized. Although the gap structures are not yet clearly known, non-trivial topology is highly anticipated.

## 3. TRS-BROKEN TOPOLOGICAL SUPERCONDUCTIVITY IN Sr$_2$RuO$_4$

In this section, I review the current situation of research on superconductivity in Sr$_2$RuO$_4$, one of the leading candidates for TRS-broken TSCs [11-14]. In particular, key experiments reported in recent years are described in the last half of this section.

### Evidence for the chiral-$p$-wave SC state
Sr$_2$RuO$_4$ has a layered perovskite structure consisting of conductive Ru-O layers and insulating Sr-O layers. Reflecting the crystal structure, this oxide exhibits Q2D conductivity with three cylinder-like Fermi surfaces labeled as α, β, and γ. The normal state of this oxide can be well explained by the multi-orbital Fermi-liquid picture [11]. This is actually rare among known unconventional superconductors, whose normal states mostly exhibit non-Fermi-liquid behavior.

The superconductivity of Sr$_2$RuO$_4$ was discovered in 1994 [50]. Soon afterwards, based on the similarity of its Landau parameters to those of $^3$He, the possibility of spin-triplet superconductivity in Sr$_2$RuO$_4$ was proposed [51]. Although the actual pairing mechanism may be more complicated than that of superfluid $^3$He, this prescient suggestion stimulated research to study the SC nature of Sr$_2$RuO$_4$, along with research to improve sample qualities.

The spin-triplet nature is directly confirmed by the NMR [52-55] and polarized neutron scatterings [56]; the spin susceptibility does not decrease in the SC state. Actually, the spin susceptibility can be even increased by ~2% in the SC state [55]. Such an increase is attributed to weak spin polarization of spin-triplet Cooper pairs assisted by strong energy dependence of the density of states [57]. The unusual spin-lattice relaxation rate $1/T_1$ using a $^{17}$O nuclear quadrupole resonance (NQR) technique at zero field [58] suggests the existence of the collective motion of the $d$-vector order parameter [59]. The recent discovery of half-quantum fluxoid states also provides strong support for the remaining spin-degree of freedom in the SC state [60].

The spontaneous time-reversal symmetry breaking was revealed by μSR measurements [61], and later by magneto-optical Kerr-effect measurements [62]. These experimental results are interpreted as the consequence of chiral-$p$-wave superconductivity expressed by the $d$-vector order parameter $\boldsymbol{d} = \Delta\boldsymbol{z}(k_x \pm ik_y)$. Here, the vector $\boldsymbol{z}$ means that the spins of Cooper pairs align perpendicular to the z axis, i.e. $S = 1$ and $S_z = 0$; and the orbital part $k_x \pm i k_y$, consisting of a complex combination of two order-parameter components $k_x$ and $k_y$, indicates an angular-momentum polarized state ($L = 1$ and $L_z = \pm 1$) with TRS breaking.

**SC phase diagram and pair breaking**

One of the most important unresolved issues regarding superconductivity of $Sr_2RuO_4$ is the unusual SC phase diagram. It has been known that the upper critical field $H_{c2}$ exhibits strong suppression when the magnetic field is applied parallel to the conductive $ab$ plane [63]. In addition, physical quantities such as magnetization, specific heat, and thermal conductivity exhibit rapid recoveries to the normal-state values just below $H_{c2}$ [64,65].

In 2013, a magnetocaloric-effect study of ultra-pure $Sr_2RuO_4$ samples revealed that the phase transition between the SC and normal states at $H_{c2}$ is actually the first-order transition [66], rather than the second-order transition for ordinary type-II superconductors (Fig. 2). In this study, a finite jump in the entropy as well as clear supercooling and/or superheating are observed. The first-order nature of the transition has been confirmed in specific-heat [67] and magnetization studies [68].

For the ordinary orbital effect, in which superconductivity is destroyed by the kinetic energy cost of supercurrents flowing around quantum vortices, the SC transition at $H_{c2}$ is expected to be second order. Thus, the observed first-order transition strongly indicates that the superconductivity is destroyed by a mechanism other than the orbital effect. This scenario is supported by the small-angle neutron-diffraction study [69], in which the intrinsic anisotropy of the coherence lengths $\xi_{ab} / \xi_c$ deduced from the vortex lattice configuration is revealed to be 60, much larger than the observed anisotropy in $H_{c2}$; i.e., $H_{c2\,//\,ab} / H_{c2\,//\,c} \sim 20$. These ratios would be equal to each other if $H_{c2}$ were purely determined by the orbital effect. The anisotropy of 60 is also deduced from analysis of magnetic torque [68].

The first-order transition resembles the Pauli pair-breaking effect occurring in spin-singlet superconductors or possibly in spin-triplet superconductors with finite spin-susceptibility reduction. Indeed, there are theoretical suggestions attributing the first order transition to the Pauli effect [70]. On the other hand, the Pauli effect is incompatible with NMR [52-55] and neutron-scattering [56] experiments: The Pauli effect should be absent in $Sr_2RuO_4$, since the characteristic field of the Pauli effect $H_P \sim 1 / (\chi_n - \chi_s)^{1/2}$ is infinitely large if the spin susceptibility in the SC state $\chi_s$ is the same as that in the normal state $\chi_n$. I should emphasize that the Knight shift of $Sr_2RuO_4$ is dominated by the Pauli spin susceptibility, as evidenced by the observed negative hyper-fine coupling at the Ru site and by the theoretical calculation [71]. Thus, the situation of $Sr_2RuO_4$ is clearly distinct from some conventional spin-singlet superconductors (such as vanadium) that do not exhibit change in the Knight shift in the SC state; the absence of the Knight-shift change in such materials is merely because the Pauli-susceptibility contribution in the Knight shift is just too small [72].

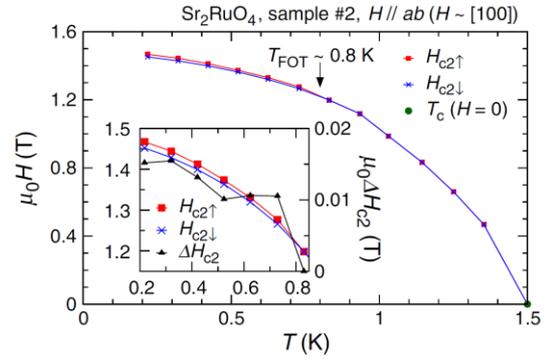

**Figure 2**. Superconducting phase diagram of $Sr_2RuO_4$ for magnetic fields parallel to the $ab$ plane, revealed by the magneto-caloric effect measurements [66]. The first-order transition was newly revealed below $T_{FOT} \sim 0.8$ K. This fact indicates the existence of an unusual pair-breaking mechanism. *Copyright (2013) American Physical Society.*

So far, there is no convincing scenario explaining the experimental results, including the observed first-order transition. One possibility is that the pairing glue is directly affected by the magnetic field.

It is worth commenting here that another candidate for chiral spin-triplet superconductivity $UPt_3$ faces a similar situation for fields oriented along the $c$ axis [63]; an unusual limitation of $H_{c2}(T)$ is seen [73] despite the absence of the Knight-shift change in the SC state [20,21]. The origin has not been clarified, as in the case for $Sr_2RuO_4$. Interestingly, $URu_2Si_2$, which is believed to be a chiral spin-*singlet* superconductor, also exhibits limited $H_{c2}(T)$ [27], but an absence of clear NMR Knight-shift change in the SC state has been reported [28]. Since this NMR measurement of $URu_2Si_2$ was performed with polycrystalline samples, it is highly necessary to perform NMR and neutron-scattering studies using high-quality single crystals.

**Enhancement of superconductivity under strain**
To understand pairing glue as well as superconducting symmetry, the study of strain-effects is of primary interest. It has been known that $Sr_2RuO_4$-Ru eutectic crystals, available under Ru-rich growth conditions, exhibit non-bulk superconductivity with $T_c$ of up to around 3.5 K (often called the "3-K phase") [74]. This enhancement has been attributed to the local strain near the interface induced by the lattice mismatch between $Sr_2RuO_4$ and Ru. Indeed, superconductivity with $T_c \sim 3.5$ K with a bulky nature can be induced by in-plane uniaxial pressure [75], and also by in-plane "bi-axial" pressure [76]. Similar Tc increases have been also reported for out-of-plane uniaxial pressure [77] but this increase may have been due to induced in-plane strain [78]. It is also revealed that lattice dislocations of $Sr_2RuO_4$ hosts superconductivity of $T_c$ of, again, around 3.5 K [79].

More recently, a uniaxial strain effect study using a novel strain-application instrument consisting of piezoelectric devices revealed that $T_c$ of $Sr_2RuO_4$ increases substantially under both compressive and tensile strains along the [100] direction [80]. This is consistent with the two-component SC order parameter. An additional interesting observation is the substantial in-plane anisotropy; the strain effect along the [110] direction is found to be much less substantial. This anisotropy is attributed to the existence of the Van Hove singularities along the [100] direction and an provides important clue for the superconducting mechanism of $Sr_2RuO_4$.

**Chiral edge currents and domains**
The chiral superconductivity is characterized by chiral edge states emerging at the surface of the superconductor. Naively, it is anticipated that the chiral edge state carries chiral edge current circulating around the edge of the superconductor. In $Sr_2RuO_4$, although the existence of unusual edge *states* is confirmed by the tunneling spectroscopy with junctions attached to surfaces perpendicular to the *ab* plane [81], observation of edge *current* has never been reported. One direct method to detect the edge current is to observe local magnetic fields or moments generated by the circulating edge current. However, according to careful investigations of surface magnetic-field distribution by using scanning Hall and SQUID sensors [82], the spontaneous magnetic field generated by the edge current is negligibly small. Detection of the magnetic moment of $Sr_2RuO_4$ microscopic rings and discs induced by the edge current have also failed [60,83].

The absence of positive experiments for the chiral edge current has been one of the fundamental challenges for the chiral-*p*-wave scenario of $Sr_2RuO_4$. It is worth noting, however, that the chiral edge current in the $^3$He-A phase, the most established chiral condensate, has not been observed either. Also, the chiral edge current has never been reported in other chiral-superconductor candidates. There may be an overlooked mechanism to suppress the anticipated chiral edge current.

Another interesting consequence of the chiral order is the possible formation of multiple domains with positive and negative chiralities. In many junction-based experiments of $Sr_2RuO_4$ performed by various groups, unusual transport properties attributable to chiral domains and domain walls have been observed [84-87]. However, direct observation of chiral domains has not been reported in $Sr_2RuO_4$. It is also worth considering the energy cost to form chiral domain walls; based on the London theory, single-domain states are revealed to be energetically more favorable than multi-domain states [88]. Thus, energetics and dynamics of the chiral domains and domain walls are still an interesting and unsettled issue.

**Toward observation of topological phenomena**
After the topological aspect of the superconductivity in $Sr_2RuO_4$ was widely recognized, a number of topological phenomena were predicted. Quantization of the thermal Hall conductivity of $Sr_2RuO_4$ in zero magnetic field [89] would provide conclusive evidence for the chiral SC state, if experimentally observed. Electrical Hall conductivity, which is not quantized, can also take a finite value at zero field [90]. However, capacitive techniques are probably required to detect such a zero-field Hall effect.

In 2D chiral-*p*-wave superconductors, it has been known that a Majorana zero mode, exhibiting non-Abelian statistics, exists in the core of a half-quantum vortex (HQV) [7]. In integer-quantum vortices, there always exist an even number of Majorana zero modes corresponding to the two-fold spin degeneracy, and these two modes interfere to lose the non-Abelian nature. Thus, half vortices, in which an odd number of Majorana zero modes exist, are required to avoid such loss of the interesting statistics. When there are several

HQVs with Majorana zero modes, adiabatic braiding of HQVs (thus zero modes) results in a change in the quantum state [7]. This change can be utilized for quantum calculations. Signatures of HQVs (strictly speaking, half-quantum fluxoids) have been observed in ultra-sensitive cantilever torque magnetometry of $Sr_2RuO_4$ micro rings [60]. In this experiment, half-height magnetization steps were found to emerge on the border of neighboring integer fluxoid states, when an additional magnetic field along the ring was applied. As a next step toward understanding the physics of HQVs and Majorana zero modes, it is desirable to detect HQVs with other techniques. Recently, magnetoresistance studies of $Sr_2RuO_4$ micro rings have been reported, but evidence for HQV has not been observed [91]. Nevertheless, such efforts are important toward experimental demonstration of non-Abelian statistics by using, e.g., the Aharonov-Casher interference as suggested in Ref. [92].

Another interesting direction is to fabricate new artificial structures toward observation of fascinating phenomena. As a promising example, a hybrid system between $Sr_2RuO_4$ and the ferromagnetic metal $SrRuO_3$ has been recently reported (Fig. 3) [93]. In a ferromagnetic metal, ordinary spin-singlet Cooper pairs are expected to be unstable due to the strong magnetic exchange field. However, spin-triplet Cooper pairs can survive deep into the ferromagnet depending on the relative direction between the Cooper pairs and the magnetization [94], the latter of which is controllable by external magnetic fields. Thus, utilization of spin degree of Cooper pairs can be achieved using this new hybrid system. Demonstration of such novel functionality would in turn provide firm evidence for the superconducting order parameter in $Sr_2RuO_4$.

## 4. TRS-PRESERVED TOPOLOGICAL SUPERCONDUCTIVITY IN $Cu_xBi_2Se_3$

In this section, I explain superconductivity in $Cu_xBi_2Se_3$, one of the leading candidates for TRS-preserved TSCs, a counterpart of the TRS-broken $Sr_2RuO_4$.

**Normal-state properties**
The mother compound, $Bi_2Se_3$, is known as a typical topological insulator. The crystal structure consists of triangular-lattice layers of Bi and Se. The five layers of the stacking order Se-Bi-Se-Bi-Se, the quintuple layer, is a unit of the crystal structure. Cu can be intercalated in the van der Waals gap between quintuple layers to make the compound a superconductor ($T_c$ ~ 3.8 K).

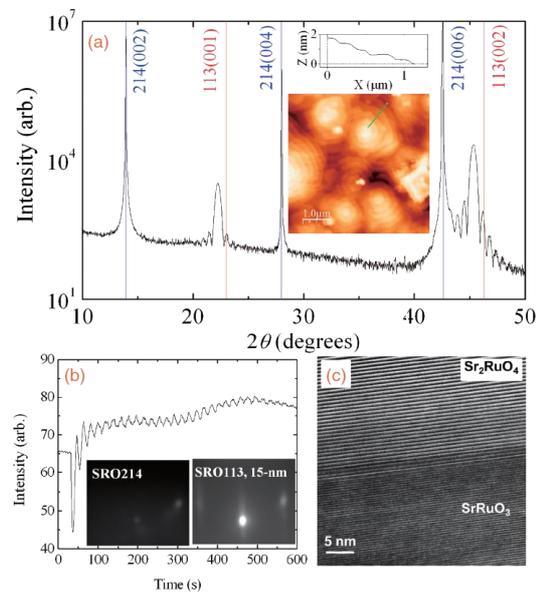

**Figure 3**. New hybrid system of $SrRuO_3$ (ferromagnetic metal) and $Sr_2RuO_4$ [93]. A thin film of $SrRuO_3$ was epitaxially grown on the *ab* surface of a $Sr_2RuO_4$ single crystal. The system exhibits ferromagnetism below ~160 K and the electrical contact between the two oxides is found to be really good. Because of these conditions, this system is suitable for study of the interference between ferromagnetism and spin-triplet superconductivity. *Copyright (2015) The Japan Society of Applied Physics.*

The normal-state electronic structure has been revealed with ARPES and quantum oscillation experiments [95]. According to these experiments, superconducting $Cu_xBi_2Se_3$ has electron carriers of $10^{20}$ cm$^{-3}$ with a cylindrical Q2D Fermi surface, rather than an ellipsoidal 3D surface.

**Superconductivity in $Cu_xBi_2Se_3$**
Superconductivity in $Cu_xBi_2Se_3$ has been reported in 2010 [30]. Point-contact spectroscopy reveals a conductance peak at zero bias voltage in the tunneling spectra, indicating the existence of unusual surface states [96]. This observation is consistent with topological superconductivity. On the other hand, the scanning tunneling microscopy (STM) on the *ab* surface reported fully-gapped *s*-wave-like spectra [97], which in fact may be *inconsistent* with the conventional pairing [98]. The temperature dependence of the specific heat suggests full-gapped superconducting states [99]. The $H_{c2}$ curve determined by resistivity reaches 5-6 T at low temperatures for $H // ab$ without any sign of the Pauli effect [100].

The possibility of odd-parity topological SC states was proposed just after the discovery of superconductivity of $Cu_xBi_2Se_3$ [101]. Owing to the multi-orbital nature as well as the strong spin-orbit coupling, topological superconductivity can be realized even with simple pairing channels. Among the proposed superconducting states named $\Delta_1$, $\Delta_2$, $\Delta_3$, $\Delta_{4x}$, and $\Delta_{4y}$, all except for $\Delta_1$ are odd-parity superconductivity and topologically non-trivial [102,101,3,4]. Various SC properties are calculated for these predicted states. The observed zero-bias peak in the tunneling spectra is revealed to be consistent with the $\Delta_3$, $\Delta_{4x}$, and $\Delta_{4y}$ states [103]. The SC states $\Delta_{4x}$ and $\Delta_{4y}$, possessing two gap nodes or minima along the conduction plane, are of particular interest since the gap amplitudes break the trigonal rotational symmetry of the lattice [104]. In analogy to the nematic liquid-crystal phases spontaneously breaking rotational symmetry, the $\Delta_{4x}$ and $\Delta_{4y}$ states can be called "nematic superconductivity" [105].

Since 2015, several new reports on bulk experiments indicating anomalous SC features have been posted to the arXiv server. In particular, an NMR study using single crystalline samples revealed spin-triplet superconductivity and the spin-rotational-symmetry breaking in the SC state [106]. Also, other dopants (such as Sr, Nb, and Tl) driving $Bi_2Se_3$ to superconductors have been reported [107-110]. These new experiments will certainly trigger further research to reveal the true nature of superconductivity in this compound.

## 5. SUMMARY

Currently known candidate materials for bulk topological superconductivity is reviewed in this article, with some more in-depth analysis for the two leading candidates $Sr_2RuO_4$ and $Cu_xBi_2Se_3$. The recent extension of the concept of topology to nodal superconductivity allows expansion of the candidate list toward various long-studied unconventional superconductors, including cuprates, heavy fermions, and organics. Experimental efforts toward observation of topological phenomena in both gapped and gapless topological superconductors are important and interesting tasks for the near future.

## ACKNOWLEDGEMENTS

I would like to thank A. Furusaki, Y. Yanase, S. Kittaka, S. Kasahara, K. Ishida, Y. Ando, Y. Nagai, and Y. Maeno for their insights and for our productive discussions. This work, as well as some studies explained in the text, have been supported by the Grants-in-Aids for Scientific Research on Innovative Areas on Topological Quantum Phenomena (KAKENHI 22103002) and Topological Materials Science (KAKENHI 15H05852), and the Grants-in-Aids for Scientific Research (KAKENHI 21740253, 23540407, 23110715, and 26287078) from the Japan Society for the Promotion of Science (JSPS).

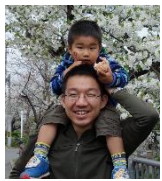

**Shingo Yonezawa** is an assistant professor at the Graduate School of Science, Kyoto University. He received a D.Sc. from Kyoto University for his study on the properties of quasi-one-dimensional organic superconductors. Since 2008, he has been working at the Graduate School of Science in Kyoto University. His main research interest is to discover interesting phenomena related to unconventional and topological superconductivity, mainly by means of thermodynamic, magnetic, and transport techniques. He also works on various unusual electronic phenomena.